\documentclass[fp]{jpsj3}
\usepackage{bm}

\title{Magnetic Phase Diagram and Skyrmions of the Hubbard Model on the $\beta$-Mn Type Lattice}

\author{Yoshiro Kakehashi\thanks{yok@sci.u-ryukyu.ac.jp, to be published in J. Phys. Soc. Jpn.}}
\inst{Department of Physics and Earth Sciences, Faculty of Science,\\
University of the Ryukyus, \\
1 Senbaru, Nishihara, Okinawa, 903-0213, Japan 
\vspace{10mm} } 

\abst{
Magnetic phase diagram for the Hubbard model on the $\beta$-Mn type lattice has been calculated as a function of the Coulomb interaction energy parameter $U$ and the electron number per atom $n$ by using the generalized Hartree-Fock approximation combined with the recursion method for electronic-structure calculations.  The ferromagnetic state, the ferrimagnetic state, and the helimagnetic state as well as the 3$Q$ multiple spin density waves (3QMSDW) states have been obtained on the $U$-$n$ plane.  Their detailed structures are examined with use of the Fourier analysis.  It is shown that the calculated phase diagram and the Dzyaloshinskii-Moriya interaction elucidate the magnetic interactions for the itinerant-electron skyrmions in transition metal alloys and compounds on the $\beta$-Mn type lattice.
}

\begin{document}
\maketitle

\section{Introduction}

\begin{figure}[htbp]
\begin{center}
\includegraphics[width=8cm]{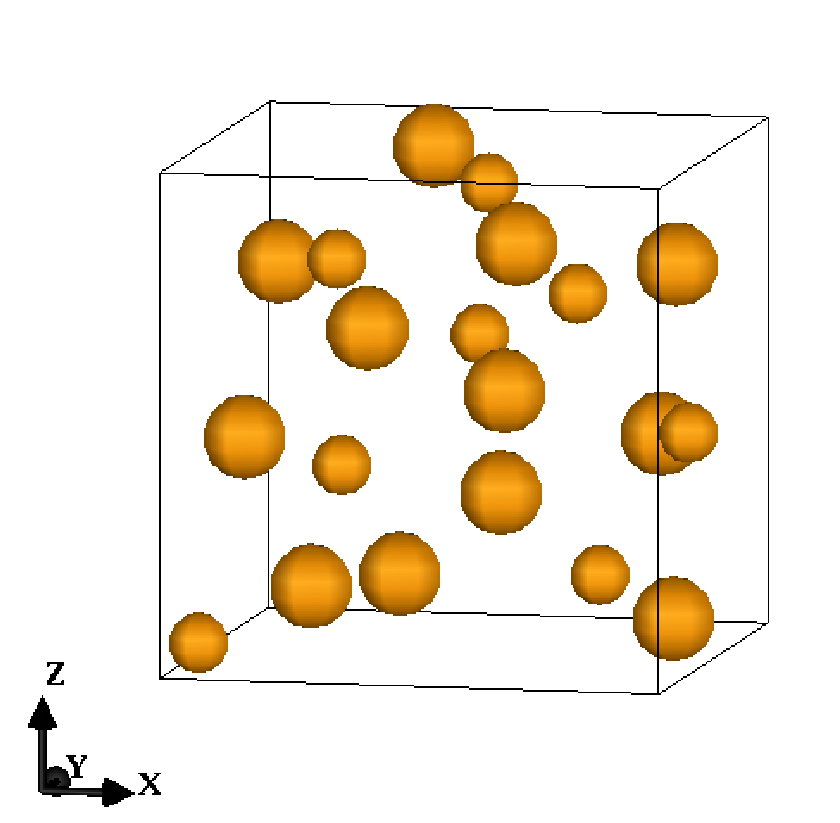}
\end{center}
\vspace{0cm}
\caption{(Color online) $\beta$-Mn type crystal structure.
There are 20 atoms in a unit cell.  Small (Large) spheres indicate the site I (II) atoms~\cite{shoe78}.
Their atomic positions are listed in Table I.
}
\label{bmnstruc}
\end{figure}

Magnetic skyrmions (Sk) are a vortex-type magnetic structure in which the magnetic moments being perpendicular to the vortex plane at the core center incline spirally away from the center and point in the opposite direction on the periphery~\cite{skyr61,bogd89,bogd94,bogd01,ross06} .
Because of a new type of topologically protected system and their potential applications to nano-size magnetic devices, magnetic skyrmions and related materials have been much investigated in the past decade~\cite{fert17,ever18}.
Although these magnetic Sk and Sk lattices have been found typically in the B20-type transition metal alloys and compounds such as MnSi~\cite{muhl09} and FeGe~\cite{yu11}, a new class of Sk metals and alloys have recently been found in the systems with the $\beta$-Mn type lattice~\cite{toku14,karu20,yu18,qian20}.
The $\beta$-Mn type crystal structure with the space group of P4${}_{1}$32 has 20 atoms in a unit cell~\cite{shoe78}.  There are two inequivalent sites called site I (8c) and site II (12d). 
Among 20 atoms, 8 atoms are on site I and 12 atoms are on site II as shown in Fig. \ref{bmnstruc}.

The $\beta$-Mn type chiral magnet Co${}_{10}$Zn${}_{10}$ showing the helical structure at the ground state ($T=0$) has been found to form the triangular Sk lattice near the critical temperature $T_{\rm C}$ of the helical state under the magnetic field~\cite{toku14}.
Co${}_{x}$Zn${}_{y}$Mn${}_{z}$ ($x+y+z=20$) alloys with $0 < z <3$ are also reported to show 
the helical structure at $T=0$ and 
the triangular Sk lattice near $T_{\rm C}$~\cite{toku14,karu20}, while a half-Sk lattice is found in Co${}_{8}$Zn${}_{9}$Mn${}_{3}$ alloy~\cite{yu18}.
Co${}_{x}$Zn${}_{y}$Mn${}_{z}$ ($x+y+z=20$) with more Mn concentration ($3 < z < 7$), on the other hand, show the reentrant spin glass state (RSG) at the ground state~\cite{karu20},  and form the triangular Sk lattice near $T_{\rm C}$ in general.
In the $\beta$-Mn type Fe${}_{2-x}$Pd${}_{x}$Mo${}_{3}$N thin film with the ferromagnetic ground state~\cite{qian20}, two types of the Sk phases are reported below 350K: the low-temperature Sk-I and the high-temperature Sk-II.
%
%
%
%

The Sk in these alloys and compounds have been considered to be caused by
a competition between the magnetic interactions in their systems.
Since the $\beta$-Mn type structure has no inversion symmetry, the Dzyaloshinskii-Moriya (DM) interactions~\cite{dzya58,mori60} play a significant role in the Sk systems.
The helical structure and Sk in Co${}_{10}$Zn${}_{10}$ are considered to be stabilized by a competition between the ferromagnetic (F) exchange interactions due to itinerant $d$ electrons and the DM interaction~\cite{toku14}.
The helical or RSG ground state and the Sk in Co${}_{x}$Zn${}_{y}$Mn${}_{z}$ ($x+y+z=20$) alloys are considered to be caused by the competition between the ferromagnetic  interaction and the antiferromagnetic (AF) interaction~\cite{toku14,karu20}. 
On the other hand, the Sk in the Fe${}_{2-x}$Pd${}_{x}$Mo${}_{3}$N system is considered to be created by the competition between the ferromagnetic interactions and the DM interactions, since FeMo${}_{3}$N compound shows the ferromagnetism~\cite{qian20}.

Although above-mentioned scenarios on the formation of the itinerant-electron Sk seem to be plausible theoretically,  systematic and self-consistent determination of the magnetic structure of  the Sk systems on the $\beta$-Mn type lattice have not yet been made in the model calculation.  Recently, there are some electronic-structure calculations that successfully evaluated the electron-filling dependence of the DM interaction in this material~\cite{karu18,luo19}.
In the present paper, we consider the Hubbard model~\cite{hub63,hub64,gutz63,gutz64,gutz65} on the $\beta$-Mn type lattice and calculate the magnetic phase diagram on the basis of the Generalized Hartree-Fock approximation (GHF) and the recursion method for electronic structure calculations~\cite{hay75,heine80}.  The calculations allow us to understand the magnetic structures and magnetic interactions of the Sk system on the $\beta$-Mn type lattice.
The same approach has been applied to the face-centred cubic lattice system and the metallic Sk crystals have been found~\cite{kake20}.
More recently, Kobayashi and Hayami also found the metallic Sk on the triangular lattice on the basis of the Hubbard model~\cite{koba22}. Makuda and Hotta found various types of Sk at half filling on the same lattice using the Hubbard model with the Rashba spin-orbit coupling~\cite{maku24}. 

We obtained the magnetic phase diagram consisting of the ferromagnetic state, the ferrimagnetic state, and the helimagnetic state on the plane of the Coulomb interaction energy parameter $U$ and the electron number per site $n$.
Moreover, we found the triple $Q$ multiple spin density waves states (3QMSDW) as well as the partially-ordered spin density wave state (partially-ordered SDW).
On the basis of the phase diagram we clarify the magnetic interactions in the itinerant-electron Sk systems on the $\beta$-Mn type lattice, which are consistent with those obtained from the experimental point of view.

In the following section, we briefly explain the model and method of numerical calculations.  In Sect. 3, we present the numerical results of the magnetic structures and elucidate their stability on the basis of their electronic structure.  Since the $\beta$-Mn type lattice consists of 20 atoms in a unit cell, calculated magnetic structures are site-dependent and complex even in the unit cell.  We therefore present analytic expressions of the magnetic structures as well as the application visualization system (AVS) images.  On the basis of the calculated phase diagram, we elucidate the magnetic interactions of the Sk systems on the $\beta$-Mn lattice.
In the last section, we summarize the numerical results obtained in the present work.

\section{Model and Method of Calculations}

     We consider the Hubbard model~\cite{hub63,hub64,gutz63,gutz64,gutz65} on the $\beta$-Mn type lattice. 
\begin{eqnarray}
\hat{H} = \sum_{i \sigma} \epsilon_{i} \, n_{i\sigma} 
+ \sum_{i j \sigma} t_{i j} \, a_{i \sigma}^{\dagger} a_{j \sigma}
+ \sum_{i} U_{i} \, n_{i \uparrow}n_{i \downarrow} \ .
\label{hub-h}
\end{eqnarray}
Here  $\epsilon_{i}$,  $t_{i j}$, and $U_{i}$ are the atomic level on site $i$, the transfer integral between sites $i$ and $j$, and the intra-atomic Coulomb interaction energy parameter on site $i$, respectively.  $a_{i \sigma}^{\dagger} (a_{i \sigma})$ is the creation 
(annihilation) operator for an electron with spin $\sigma$ on site $i$,
and $n_{i\sigma}=a_{i \sigma}^{\dagger} a_{i \sigma}$ is the number
operator of electron on site $i$ for spin $\sigma$. We have omitted the Dzyaloshinskii-Moriya (DM) interaction in the present calculations.

The $\beta$-Mn type lattice has 20 atoms in a unit cell with two inequivalent sites, site I and site II, as have been introduced in the last section.  In the present calculations, we adopted the ideal $\beta$-Mn type lattice~\cite{shoe78,naka97}.  The atomic positions in the unit cell are  listed in Table I.  We adopt the same atomic level (Coulomb interaction energy) for both sites I and II; $\epsilon_{i} = \epsilon_{0} = 0$ ($U_{i} = U$).
Moreover, we take into account the transfer integrals $t_{k}$ up to the 6-th nearest neighbors (NN): $t_{1}=-1.0$ and $t_{k} = -0.5$ ($k = 2 \sim 6$) in unit of the NN transfer integral ($|t_{1}| \equiv 1$).  The choice of parameters $t_{k}$ is based on Heine's law for 
3$d$ transition metals; $t_{ij} \propto R^{-5}$, where $R$ is the inter-atomic distance between sites $i$ and $j$, as has been discussed in our previous paper~\cite{kake22}.
%
%
\begin{table}[tbh]
\caption{Atomic sites  $k$ and atomic positions $(x, y, z)$ in the ideal $\beta$-Mn type lattice~\cite{shoe78,naka97}.
8 atoms $(k=1-8)$ are on the crystallographic equivalent site called site I, and the remaining 12 atoms $(k=9-20)$ are on site II.  Atomic positions are given in unit of lattice parameter $a$.
\vspace{5mm} }
\label{table-crd}
\begin{tabular}{cccc}
\hline
$k$  & x & y & z  \\ \hline
\vspace*{1mm}
 1 & 0.0636 & 0.0636 & 0.0636 \\
 2 & 0.3136 & 0.1864 & 0.8136 \\
 3 & 0.8136 & 0.3136 & 0.1864 \\
 4 & 0.1864 & 0.8136 & 0.3136 \\
 5 & 0.6864 & 0.6864 & 0.6864 \\
 6 & 0.5636 & 0.4364 & 0.9364 \\
 7 & 0.4364 & 0.9364 & 0.5636 \\
 8 & 0.9364 & 0.5636 & 0.4364 \\ \hline
 9 & 0.7035 & 0.0465 & 0.8750 \\
10 & 0.8750 & 0.7035 & 0.0465 \\
11 & 0.0465 & 0.8750 & 0.7035 \\
12 & 0.3750 & 0.7965 & 0.9535 \\
13 & 0.9535 & 0.3750 & 0.7965 \\
14 & 0.7965 & 0.9535 & 0.3750 \\
15 & 0.4535 & 0.1250 & 0.2035 \\
16 & 0.2035 & 0.4535 & 0.1250 \\
17 & 0.1250 & 0.2035 & 0.4535 \\
18 & 0.5465 & 0.6250 & 0.2965 \\
19 & 0.2965 & 0.5465 & 0.6250 \\
20 & 0.6250 & 0.2965 & 0.5465 \\
\hline 
\end{tabular}
\end{table}
%
%

In order to calculate the magnetic structures on the $\beta$-Mn lattice, we apply the Generalized Hartree-Fock (GHF) approximation for the Hubbard model (\ref{hub-h}).  In this approximation, we introduce first locally rotated coordinates on each site, and adopt the Hartree-Fock approximation.  We then obtain the GHF Hamiltonian as follows.
\begin{eqnarray}
\hat{H} = \sum_{i \alpha j \gamma} a_{i \alpha}^{\dagger} H_{i \alpha j \gamma} a_{j \gamma}
- \sum_{i} \frac{1}{4} \, U \, ( \langle n_{i} \rangle^2 - \langle \boldsymbol{m}_{i} \rangle^2 ) \ .
\label{ghf-h}
\end{eqnarray}
The one-electron Hamiltonian matrix element $H_{i \alpha j \gamma}$ is given by
\begin{eqnarray}
H_{i \alpha j \gamma} = 
\Big[ \big( \epsilon_{0} - \mu + \frac{1}{2} U \langle n_{i} \rangle \big) \delta_{\alpha \gamma} - 
\frac{1}{2} U \langle \boldsymbol{m}_{i} \rangle \cdot (\boldsymbol{\sigma})_{\alpha \gamma}
\Big] \delta_{ij} + t_{ij} \delta_{\alpha\gamma} (1-\delta_{ij}) .
\label{hmatrix}
\end{eqnarray}
Here we introduced the chemical potential $\mu$. $\boldsymbol{\sigma} = (\sigma_{x}, \sigma_{y}, \sigma_{z})$ are the Pauli spin matrices.  $\langle n_{i} \rangle$ and 
$\langle \boldsymbol{m}_{i} \rangle$ are the average local charge and magnetic moment on site $i$, respectively.

The local charge and magnetic moment in Eqs. (\ref{ghf-h}) and (\ref{hmatrix}) are given in the GHF as follows.
\begin{eqnarray}
\langle n_{i} \rangle = \int d\omega f(\omega) \sum_{\alpha}
\rho_{i\alpha\alpha}(\omega) ,
\label{chargeni}
\end{eqnarray}
\begin{eqnarray}
\langle \boldsymbol{m}_{i} \rangle = 
\sum_{\alpha\gamma} (\boldsymbol{\sigma})_{\alpha\gamma} 
\int d\omega f(\omega)
\rho_{i\gamma\alpha}(\omega) .
\label{spinmi}
\end{eqnarray}
Here $f(\omega)$ is the Fermi distribution function at zero temperature.
$\rho_{i\alpha\gamma}(\omega)$ is the local density of states (DOS) for one electron Hamiltonian (\ref{hmatrix}).
\begin{eqnarray}
\rho_{i\alpha\gamma}(\omega) = \sum_{\kappa} \langle i | \kappa \rangle_{\alpha} 
\delta(\omega - \epsilon_{\kappa}) \langle \kappa | i \rangle_{\gamma} \ , 
\label{rhoag}
\end{eqnarray}
where $\epsilon_{\kappa}$ and $\langle i | \kappa \rangle_{\alpha}$
are the eigenvalue and eigenvector for the one-electron Hamiltonian  (\ref{hmatrix}), respectively.

Equations  (\ref{hmatrix}),  (\ref{chargeni}), (\ref{spinmi}), and (\ref{rhoag}) form the self-consistent equations to obtain the local charges $\{ \langle n_{i} \rangle \}$ and magnetic moments $\{ \langle \boldsymbol{m}_{i} \rangle \}$.

The ground-state energy is given by 
\begin{eqnarray}
E = \mu N + \int d\omega f(\omega) \omega \rho(\omega) 
- \sum_{i} \frac{1}{4} U ( \langle n_{i} \rangle^2 - \langle \boldsymbol{m}_{i} \rangle^2 ) \ .
\label{gener}
\end{eqnarray}
Here $N$ is the total electron number and $\rho(\omega)$ is the total DOS given by $\rho(\omega)=\sum_{i\alpha} \rho_{i\alpha\alpha}(\omega)$.

In the numerical calculations, we adopted the self-consistent equation (\ref{chargeni}) in the large $U$ limit~\cite{hub81}:
\begin{eqnarray}
n_{i} = \int d\omega f(\omega) \sum_{\alpha}
\rho_{i\alpha\alpha}(\omega) \ .
\label{approxni}
\end{eqnarray}
Here $n_{i}$ $(=n)$ at the left-hand-side is the local charge on site $i$ in the nonmagnetic state and $\rho_{i\alpha\alpha}(\omega)$ at the right-hand-side is the local DOS for the Hamiltonian (\ref{hmatrix}) in which the charge potentials $\epsilon_{0} - \mu + U \langle n_{i} \rangle/2$ have been replaced by $\tilde{\epsilon}_{i}$.  The latter are determined so as to satisfy Eq. (\ref{approxni}) for a given set of $\{ \langle \boldsymbol{m}_{i} \rangle \}$. 

It should be noted that the DOS  (\ref{rhoag}) are given by the on-site one electron Green function  
$G_{i\alpha i\gamma}(z) = [(z-H)^{-1}]_{i\alpha i\gamma}$ as follows.
\begin{eqnarray}
\rho_{i\alpha\gamma}(\omega) = \frac{(-)}{\pi} {\rm Im} \, G_{i\alpha i\gamma}(z) \ .
\label{rhogag}
\end{eqnarray}
Here $z = \omega + i\delta$, $\delta$ being a positive definite infinitesimal number.

In order to solve the self-consistent equations for arbitrary magnetic structure, we calculated the Green function using the recursion method~\cite{hay75,heine80}.
In this method, we transform the Hamiltonian (\ref{hmatrix}) into a tridiagonal matrix using  a recursive unitary transformation.  Then, the diagonal Green function $G_{i\alpha i\alpha}(z)$ for the calculations of $\langle n_{i} \rangle$ and $\langle m_{iz} \rangle$ is expressed by a continued fraction as follows.
\begin{eqnarray}
G_{i\alpha i\alpha}(z) = 
\cfrac{1}{z - a_{1} - 
\cfrac{|b_{1}|^{2}}
{z - a_{2} -
\cfrac{|b_{2}|^{2}} 
{\ldots \cfrac{\ddots}
{\ldots
- \cfrac{|b_{n-1}|^{2}}
{z - a_{n} - T_{n}(z)} 
}}}} \ .
\label{gcont}
\end{eqnarray}
We obtain the recursion coefficients $\{ a_{m}, b_{m} \}$ up to the $n$-th level numerically from the Hamiltonian matrix elements using the recursion algorithm.
We approximate the higher-order coefficients in the terminator $T_{n}(z)$ by their asymptotic values $a_{\infty}$, $b_{\infty}$, so that we obtain an approximate form of $T_{n}(z)$ as follows.
\begin{eqnarray}
T_{n}(z) \approx T_{\infty}(z) = \frac{1}{2} \Big(
z - a_{\infty} - \sqrt{(z - a_{\infty})^{2} - 4 |b_{\infty}|^{2}}
\Big) \ .
\label{tinfty}
\end{eqnarray}

For the calculations of $\langle m_{ix} \rangle$ ($\langle m_{iy} \rangle$), we need off-diagonal Green functions $G_{i\uparrow i\downarrow}+G_{i\downarrow i\uparrow}$ ($i (G_{i\uparrow i\downarrow}-G_{i\downarrow i\uparrow})$).  In this case, we introduce the basis set $|i1 \rangle$, $|i2 \rangle$ ($|i3 \rangle$, $|i4 \rangle$) which diagonalize 
$\sigma_{x}$ ($\sigma_{y}$), and calculate 
$G_{i\uparrow i\downarrow}+G_{i\downarrow i\uparrow}$ 
($i (G_{i\uparrow i\downarrow}-G_{i\downarrow i\uparrow})$) using the diagonal matrices 
$G_{i1 i1}$ and $G_{i2 i2}$ ($G_{i3 i3}$ and $G_{i4 i4}$) in the same way~\cite{kake13}.

\section{Numerical Results}

In the numerical calculations, we adopted a large cubic cluster consisting of $10 \times 10 \times 10$ $\beta$-Mn unit cells on a computer, which is surrounded by 26 cubic clusters with the same size and the same magnetic structure.  We calculated the recursion coefficients $\{ a_{m}, b_{m} \}$ up to the 20th level for 20000 atoms in the central $10 \times 10 \times 10$ cluster for given magnetic structure and charge potential, to solve the self-consistent equations (\ref{spinmi}) and (\ref{approxni}).
%
%
\begin{figure}[htbp]
\begin{center}
\vspace{2cm}
\includegraphics[width=10cm]{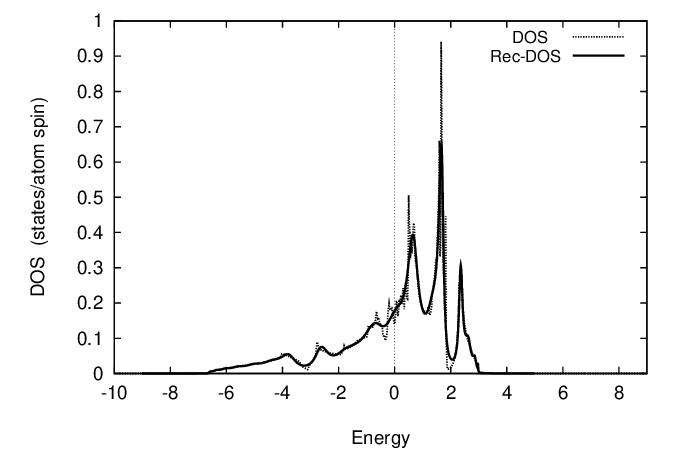}
\end{center}
\vspace{-2cm}
\caption{Non-interacting density of states per atom per spin for Hamiltonian (\ref{hmatrix}) calculated by the recursion method up to the recursion level $20$ (Rec-DOS: solid curve).
The result is compared with that is obtained by the standard $k$-integration method (DOS: dotted curve).
The energy is measured here and hereafter in unit of the nearest-neighbor transfer integral ($|t_{1}| \equiv 1$).
}
\label{recdos0}
\end{figure}
%
%
Figure \ref{recdos0} shows the noninteracting DOS calculated with use of the recursion method.  The result reproduces well the DOS obtained by the usual $k$-integration method.

We considered in the present work the ferromagnetic, paramagnetic, and helimagnetic states in order to clarify the basic property of magnetic interactions and magnetic phase diagram on the $\beta$-Mn type lattice.

\begin{figure}[htbp]
\begin{center}
\vspace{2cm}
\includegraphics[width=11cm]{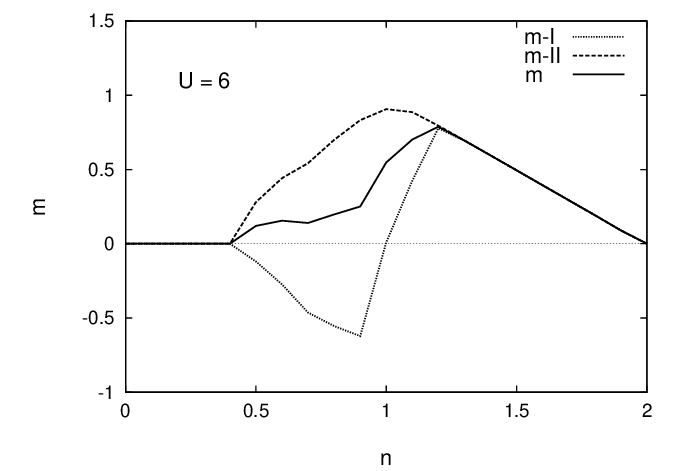}
\end{center}
\vspace{-2cm}
\caption{
Magnetization $m$ vs electron number $n$ curve at the Coulomb interaction energy $U=6$.  Solid curve : magnetization per atom ($m$), dotted curve : magnetization on site I ($m$-I), dashed curve : magnetization on site II ($m$-II).  
}
\label{mvsnu6}
\end{figure}

We performed first the self-consistent collinear calculations in which we start from the ferromagnetic structure.  Figure \ref{mvsnu6} shows a typical example of magnetization vs electron number curve calculated at the Coulomb interaction energy parameter $U=6$.  With decreasing electron number per atom $n$ from $n=2$, the magnetization per atom linearly increases first.  At $n=1.2$, the strong ferromagnetism collapses and the magnetic moments on site I ($m$-I) start to decrease with decreasing electron number.  They vanish once at $n \approx 1.0$, while the magnetic moments on site II ($m$-II) still point up there, so that the partially ordered ferromagnetic state is realized in the vicinity of $n=1.0$.
%
%
%
%
Below $n=1.0$, the magnetic moments $m$-I become anti-parallel to the magnetic moments $m$-II, so that the ferrimagnetic structure is realized.
In the region $n < 0.9$, the magnitudes of $m$-I and $m$-II decrease with decreasing $n$, and 
the ferrimagnetic state disappears below $n=0.4$.
%
%
\begin{figure}[htbp]
\begin{center}
\vspace{2cm}
\includegraphics[width=10cm]{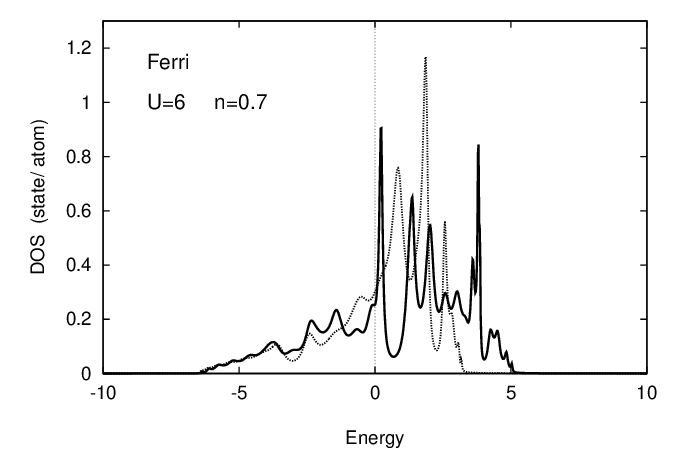}
\end{center}
\vspace{-2cm}
\caption{
Calculated densities of states per atom (DOS) in the ferrimagnetic state (solid curve) and the paramagnetic state (dotted curve) at $n=0.7$ and $U=6$.  The energy is measured from the Fermi level.
}
\label{ferridosu6}
\end{figure}
%
%

Stability of the ferrimagnetic state may be understood from the DOS as shown in Fig. \ref{ferridosu6}.  Though the ferrimagnetic structure creates a deep valley in the middle of the DOS and a sharp peak just above the Fermi level $\epsilon_{\rm F}$, it also produces a small dip below $\epsilon_{\rm F}$ around the energy $\epsilon=-1.0$ and moderate peaks around $\epsilon=-2.0$.  These dip and peaks below $\epsilon_{\rm F}$ yield the band energy gain of the ferrimagnetic state.

\begin{figure}[htbp]
\begin{center}
\vspace{2cm}
\includegraphics[width=10cm]{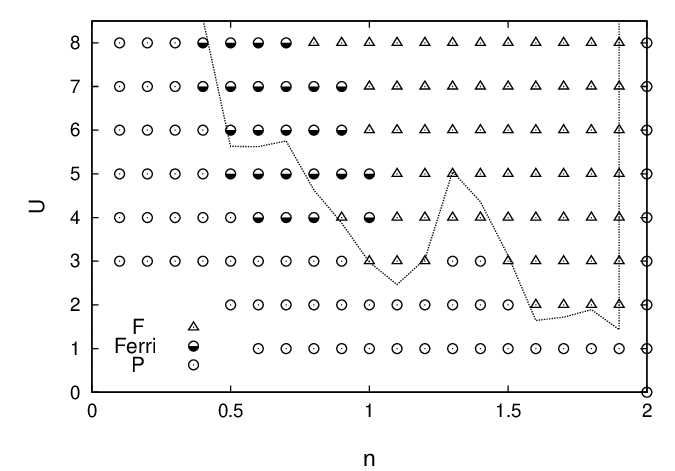}
\end{center}
\vspace{-2cm}
\caption{
Calculated $U$-$n$ phase diagram for the ferro- (F: open triangles), ferri- (Ferri: half closed circles), and para- (P: open circles) magnetic states.  Dotted line shows the Stoner boundary defined by $U_{c} = 1/\rho_{\rm II}(0)$, where $\rho_{\rm II}(0)$ denotes the local DOS on site II.
}
\label{unpdcoll}
\end{figure}

Repeating the same calculations varying $n$ and $U$, we obtained the $U$-$n$ phase diagram for the collinear states as shown in Fig. \ref{unpdcoll}.
The ferromagnetic state is stabilized in the region of $n \gtrsim 1$.  The boundary between the ferro- and para-magnetic states is basically determined by a ``Stoner'' condition for the DOS on site II, {\it i.e.}, $U_{c} = 1/\rho_{\rm II}(0)$, where $\rho_{\rm II}(0)$ denotes the local DOS on site II per atom per spin at the Fermi level.

The ferrimagnetic state appears between the ferro- and para-magnetic states above the Coulomb interaction strength $U > 3$.
It is stabilized by the short-range antiferromagnetic interactions between the magnetic moments on site I and those on site II.
We note that the partially ordered ferromagnetic states appear in general on the boundary between the ferri- and ferro-magnetic states as we have discussed in Fig. \ref{mvsnu6}.
%
%
%
%

We also performed self-consistent calculations of the helical state to clarify the  noncollinear magnetic states on the $\beta$-Mn type lattice due to competing interactions.  In the calculations, we assumed the wave vector perpendicular to the (001) helical plane, {\it i.e.}, $\bm{Q} = (0, 0, q)2\pi/a$.  Here $q$ is the wave number and $a$ is the lattice parameter.
At each point of $(n, U)$, we performed the self-consistent calculations varying the wave number $q$, and determined the wave number $Q \, (=|\bm{Q}|)$ leading to the minimum energy.  After the calculations, we examined the magnetic structure using the Fourier analysis.
%
%
\begin{figure}[htbp]
\begin{center}
\vspace{2cm}
\includegraphics[width=10cm]{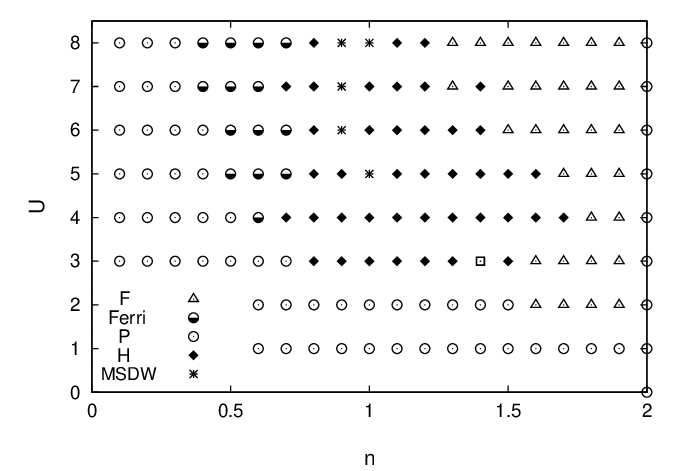}
\end{center}
\vspace{-2cm}
\caption{
Calculated $U$-$n$ phase diagram for the ferromagnetic state (F), ferrimagnetic state (Ferri), paramagnetic state (P), helimagnetic state (H), and the $3Q$ multiple spin density wave states (MSDW).  Open square at $(n, U)=(1.4, 3.0)$ indicates the partially ordered SDW state.  
}
\label{unpdncoll}
\end{figure}
%
%

\begin{figure}[htbp]
\begin{center}
\vspace{2cm}
\includegraphics[width=10cm]{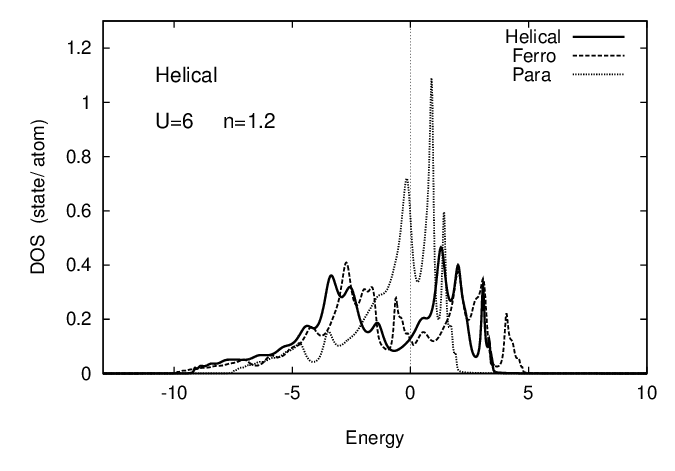}
\end{center}
\vspace{-2cm}
\caption{
Calculated DOS in the helimagnetic state (solid curve), the ferromagnetic state (dashed curve), and the paramagnetic state (dotted curve) at $(n, U)=(1.2, 6)$.
}
\label{dosh}
\end{figure}

We present the calculated phase diagram in Fig. \ref{unpdncoll}.
We find the helical state over wide range between the ferro- and ferri-magnetic states.
Figure \ref{dosh} shows the DOS in the helical state at $(n, U) = (1.2, 6)$. 
A sharp peak near the Fermi level in the paramagnetic state is much reduced by a uniform polarization in the ferromagnetic state, and the one-electron states near the Fermi level move to the lower energy region.  The DOS below the Fermi level in the ferromagnetic state further shift to the lower energy region with the helical polarization.  This causes a large band energy gain stabilizing the helical state.

%
%
%
%

\begin{figure}[htbp]
\begin{center}
\vspace{2cm}
\includegraphics[width=10cm]{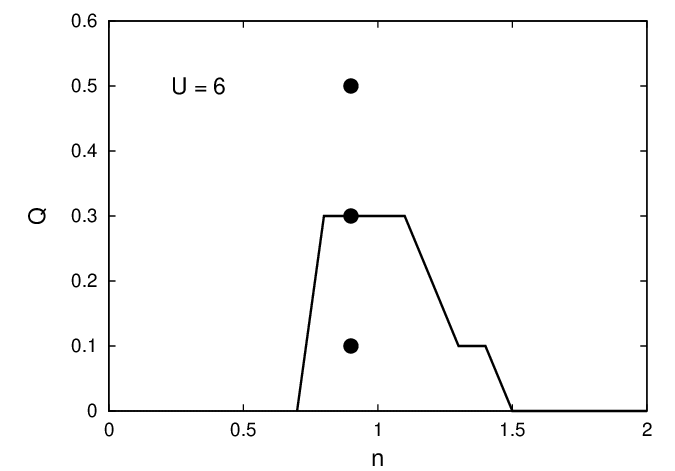}
\end{center}
\vspace{-2cm}
\caption{
Wave number $Q$ vs electron number $n$ curve at $U=6$.
$Q$ are measured in unit of $2\pi/a$, $a$ being the lattice parameter.
Closed circles at $n=0.9$ show the $3Q$ values 0.1, 0.3, and 0.5 for the 3QMSDW.
}
\label{qn}
\end{figure}

Figure \ref{qn} shows an example of the wave number $Q$ vs $n$ curve in the helical state at $U=6$.  Here and hereafter, we measure the wave number in unit of $2\pi/a$.  
The wave number increases first with decreasing electron number $n$,
but it does not exceed 0.3.  
We verified that the wave numbers do not exceed 0.5 over the range $0 < U \leqq 8$ on the $\beta$-Mn type lattice.
At $n=0.9$, we found the $3Q$ multiple spin density waves (3QMSDW) consisting of the wave vectors $\bm{Q}=(0,0,0.1)$, $(0, 0, 0.3)$, and $(0, 0, 0.5)$.
Below $n=0.8$, the ferrimagnetic state ($Q=0$) is stabilized, and the paramagnetic state in the region $n \leqq 0.4$.
%
%
%
%

\begin{figure}[htbp]
\begin{center}
\includegraphics[width=8cm]{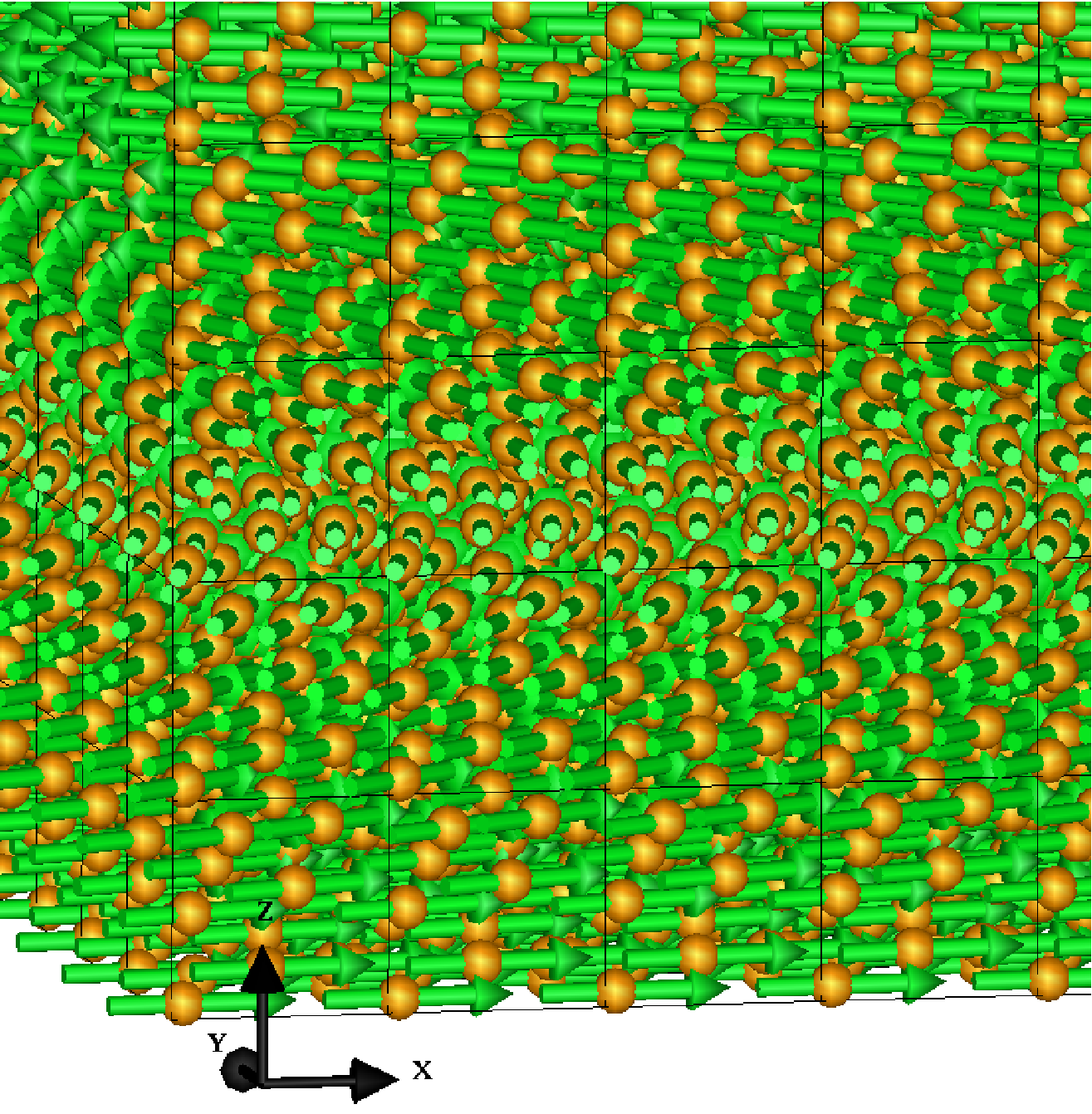}
\end{center}
\vspace{0cm}
\caption{ (Color online) 
Helical structure with the wave number $Q=0.1$ at $n=1.3$ and $U=6$. 
The magnetic moments are ferromagnetic in a crystallographic unit cell.
}
\label{hn1.3q0.1}
\end{figure}

\begin{figure}[htbp]
\begin{center}
\includegraphics[width=8cm]{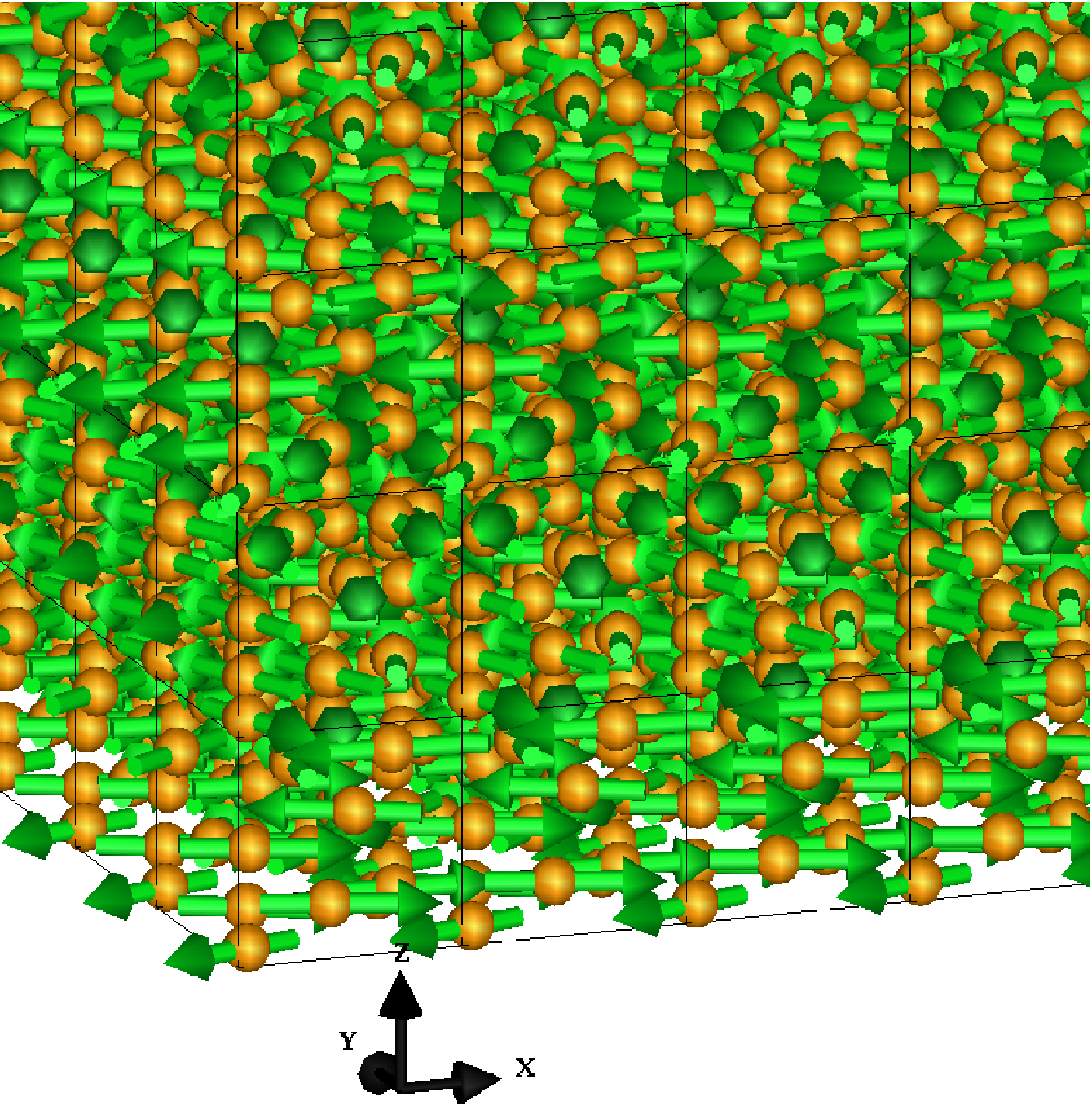}
\end{center}
\vspace{0cm}
\caption{ (Color online) 
Helical structure with the wave number $Q=0.3$ at $n=1.0$ and $U=6$.  
The magnetic moments are antiferromagnetic in a crystallographic unit cell.
}
\label{hn1.0q0.3}
\end{figure}

We note that the magnetic moments in a unit cell are ferromagnetic on the $xy$ plane in the helical structure with small $Q$ ($=0.1$), {\it e.g.}, at $n=1.3$ as shown in Fig. \ref{hn1.3q0.1}.  However with decreasing $n$, the short-range antiferromagnetic correlations develop, so that the magnetic moments in a unit cell become noncollinear at $n=1.2$ ($Q=0.2)$, and become the antiferro-magnetic at $n=1.0$ $(Q=0.3)$ as shown in Fig. \ref{hn1.0q0.3}.
%
%
%
%
%
%
%

We also found the $Q=0$ noncollinear states at $(n, U) = (0.8, 5)$, $(0.9, 5)$, $(1.2, 5)$, $(0.7, 4)$, and $(0.8 \sim 1.1, 3)$.  These states were obtained by the self-consistent calculations which start from the $q \neq 0$ helical states.
The magnetic moments on site I $(k=1-8)$ are given by
\begin{eqnarray}
\bm{m}(\bm{R}_{lk}) = m_{1} \bm{e}_{k} \ .
\label{q00851}
\end{eqnarray}
Here $\bm{R}_{lk}$ denotes the position vector of the $k$-th atom on the $l$-th unit cell and $\bm{e}_{k}$ is the unit vector showing the polarization on the $k$-th atom.
The amplitude $m_{1}$, polarization vectors $\bm{e}_{1}$ and $\bm{e}_{6}$ depend on the point $(n, U)$.  $\bm{e}_{4}$, $\bm{e}_{7}$, and $\bm{e}_{2}$ ($\bm{e}_{3}$, $\bm{e}_{8}$, and $\bm{e}_{5}$) are obtained by successive $\pi/2$ rotations of $\bm{e}_{1}$ ($\bm{e}_{6}$) around the $z$ axis.  

The magnetic moments on site II ($k = 9, 14, 16, 19$) in the $Q=0$ noncollinear states are given by
\begin{eqnarray}
\bm{m}(\bm{R}_{lk}) = m_{2} \bm{e}_{k} \ .
\label{q00852}
\end{eqnarray}
Here the amplitude $m_{2}$ depends on $(n, U)$.  $\bm{e}_{9} = (1/\sqrt{2}, -1/\sqrt{2}, 0)$. 
$\bm{e}_{16}$, $\bm{e}_{14}$, and $\bm{e}_{19}$ are obtained by successive $\pi/2$ rotations of $\bm{e}_{9}$ around the $z$ axis.  
The magnetic moments on site II ($k = 10 \sim 13, 15, 17, 18, 20$) are given by
\begin{eqnarray}
\bm{m}(\bm{R}_{lk}) = m_{3} \bm{e}_{k} \ .
\label{q00853}
\end{eqnarray}
Here the amplitude $m_{3}$, $\bm{e}_{10}$, and $\bm{e}_{15}$ depend on $(n, U)$.  $\bm{e}_{18}$, $\bm{e}_{20}$, and $\bm{e}_{13}$ ($\bm{e}_{17}$, $\bm{e}_{11}$, and $\bm{e}_{12}$) are obtained by successive $\pi/2$ rotations of $\bm{e}_{10}$ ($\bm{e}_{15}$) around the $z$ axis.  

\begin{figure}[htbp]
\begin{center}
\includegraphics[width=8cm]{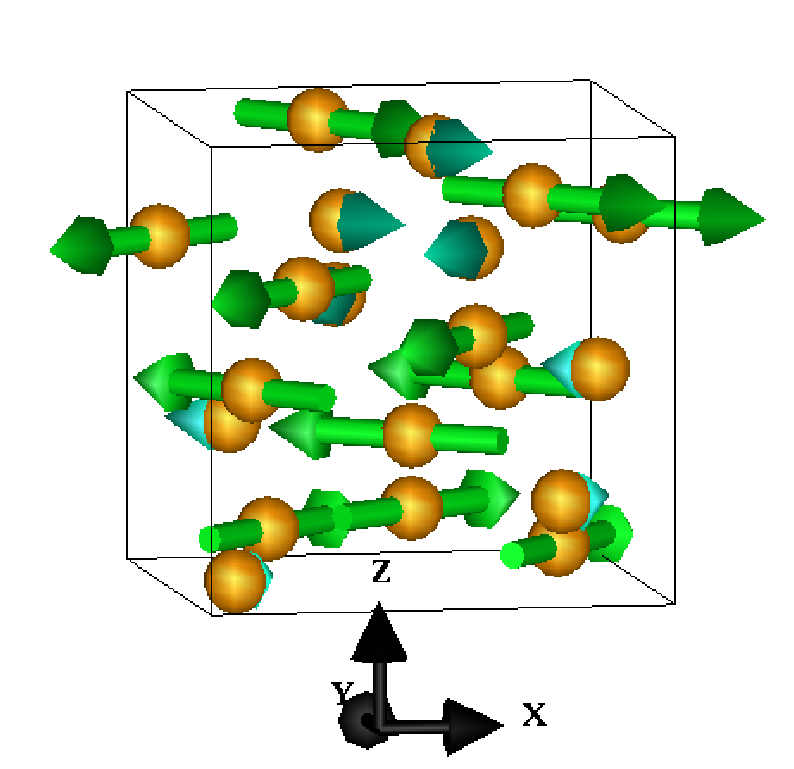}
\end{center}
\vspace{0cm}
\caption{ (Color online) 
$Q=0$ noncollinear magnetic structure in a unit cell at $(n, U) = (1.1, 3)$.
}
\label{q0nc}
\end{figure}

Figure \ref{q0nc} shows an example of the $Q=0$ noncollinear magnetic structure obtained at 
$(n, U) = (1.1, 3)$.
There the magnetic moments on site I have a small amplitude $m_{1}=0.08$, while the moments on site II have larger amplitudes $m_{2}=0.58$ and $m_{3}=0.57$, respectively.  Magnetic moments rotate along the $z$ axis.
%
%
%
%

It should be noted that the self-consistent magnetic structures do not necessarily agree with the starting ones. We obtained the triple $Q$ multiple spin density waves (3QMSDW) and the partially-ordered SDW in our calculations.

At $(n, U) = (1.4, 3)$, we obtained a partially-ordered SDW state with the wave vector $\bm{Q} = (0, 0, 0.5)$.  The magnetic moments on site I vanish.
The magnetic moments on site II are given by the site-dependent noncollinear SDW such that
\begin{eqnarray}
\bm{m}(\bm{R}_{lk}) = m_{k} \bm{e}_{k} \cos ( \bm{Q} \cdot \bm{R}_{lk} + \alpha_{k}) \ .
\label{q01431}
\end{eqnarray}
On site II ($k = 9, 14, 16, 19$), the amplitudes are given by 
$m_{k} = 0.074$, and $\bm{e}_{9} = (0.93, -0.36, 0)$, $\bm{e}_{19} = (0.36, -0.93, 0)$.  
$\bm{e}_{14}$ ($\bm{e}_{16}$) is obtained by $\pi/2$ rotation of $\bm{e}_{9}$ ($\bm{e}_{19}$) around the $z$ axis.
The site-dependent phases are given by $\alpha_{9}=0.39$, $\alpha_{16}=-\alpha_{9}$, 
$\alpha_{19}=-\alpha_{9}+\pi/2$, and  $\alpha_{14}=\alpha_{9}-\pi/2$, respectively.

On site II ($k = 10, 12, 17, 20$), we have 
$m_{k} = 0.044$, $\bm{e}_{10} = (0.85, -0.52, 0)$, and $\bm{e}_{12} = (0.85, 0.52, 0)$.  
$\bm{e}_{17}$ ($\bm{e}_{20}$) is obtained by $\pi/2$ ($-\pi/2$) rotation of $\bm{e}_{12}$ ($\bm{e}_{10}$) around $z$ axis.
The site-dependent phases are given by $\alpha_{12}=0.15$, $\alpha_{10}=-\alpha_{12}$,  
$\alpha_{17}=-\alpha_{12}+\pi/2$, and  $\alpha_{20}=\alpha_{12}-\pi/2$, respectively.

Finally, on site II ($k = 11, 13, 15, 18$),  
$m_{k} = 0.042$, $\bm{e}_{11} = (0.98, -0.21, 0)$, and $\bm{e}_{18} = (0.98, 0.21, 0)$.  
$\bm{e}_{15}$ ($\bm{e}_{13}$) is obtained by $\pi/2$ ($-\pi/2$) rotation of $\bm{e}_{11}$ ($\bm{e}_{18}$) around the $z$ axis.
The site-dependent phases are given by $\alpha_{11}=0.93$, $\alpha_{15}=\alpha_{11}$,  
and $\alpha_{13}=\alpha_{18}=-\alpha_{11}$, respectively.
%
%
\begin{figure}[htbp]
\begin{center}
\includegraphics[width=9cm]{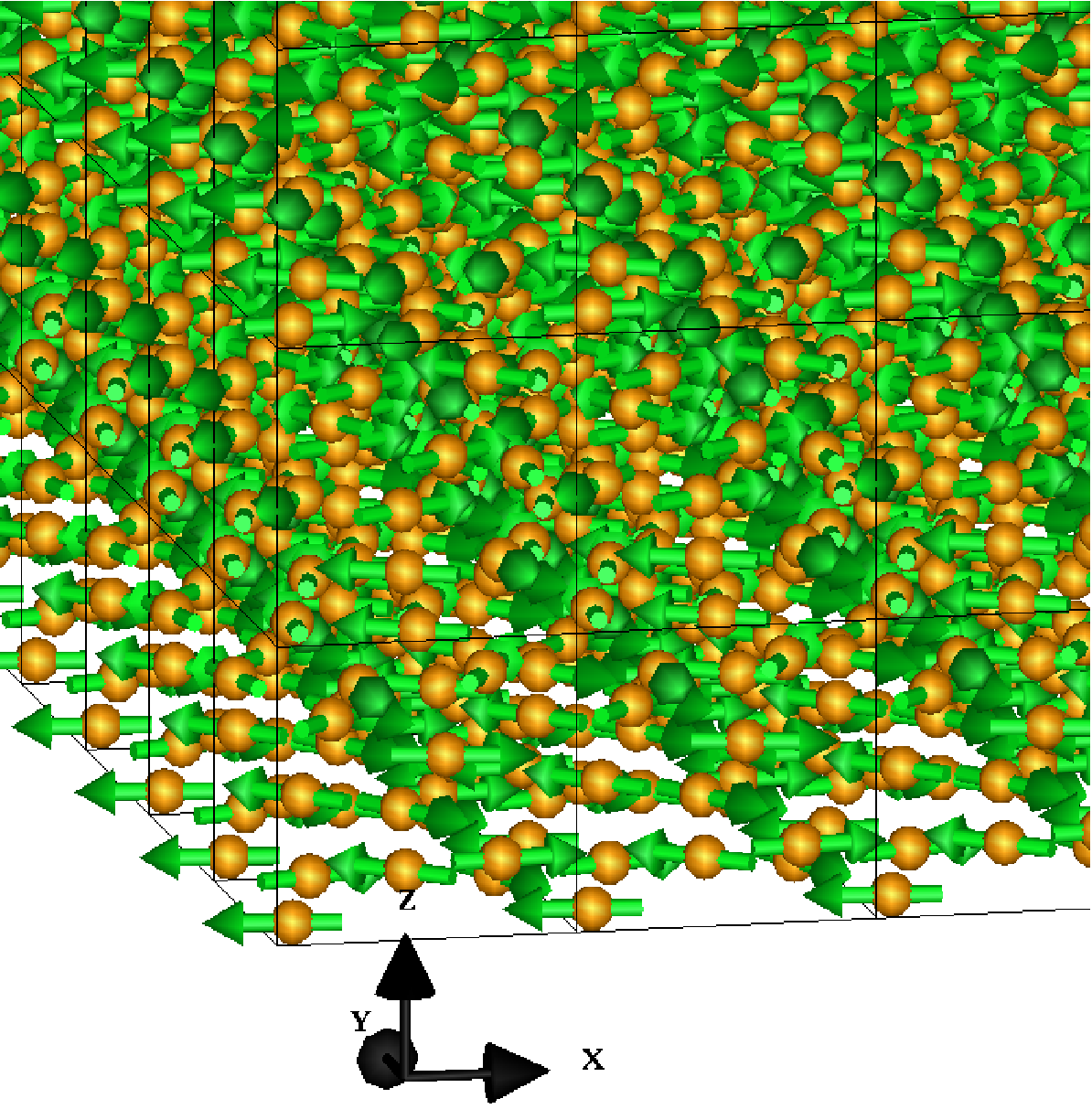}
\end{center}
\vspace{0cm}
\caption{ (Color online) 
$3Q$ multiple spin density waves (3QMSDW) obtained at $(n, U) = (1.0, 8)$.
}
\label{msdw3q}
\end{figure}
%
%

We obtained the 3QMSDW states at several points of $(n, U)$.  They show a complex noncollinear structure as shown in Fig. \ref{msdw3q}.
The 3QMSDW at $(n, U) = (1.0, 8)$ consists of the wave vectors $\bm{Q}_{0} = (0, 0, 0)$, $\bm{Q}_{2} = (0, 0, 0.2)$, and $\bm{Q}_{4} = (0, 0, 0.4)$.
The magnetic moments on site I $(k=1-8)$ are given by a site-dependent $\bm{Q}_{0}$ polarization and $2Q$ transverse SDW (2QTSDW+$Q_{0}$) such that
\begin{eqnarray}
\bm{m}(\bm{R}_{lk}) &=& m_{1} \bm{e}_{k0}  \nonumber \\
&+& m_{2} \bm{e}_{k2} \cos ( \bm{Q}_{2} \cdot \bm{R}_{lk} + \alpha_{k}) \nonumber \\
&+& m_{3} \bm{e}_{k4} \sin \bm{Q}_{4} \cdot \bm{R}_{lk}\ .
\label{3qmsdw1}
\end{eqnarray}
Here $m_{1}=0.37$, $m_{2}=0.93$, and $m_{3}=0.67$.
The polarization vectors $\bm{e}_{k0}$ in the first term 
are given by $\bm{e}_{10} = (-0.67, 0.75, 0)$, $\bm{e}_{60} = (-0.67, -0.75, 0)$.  
$\bm{e}_{70}$ ($\bm{e}_{80}$) is obtained by $\pi$ rotation of $\bm{e}_{10}$ ($\bm{e}_{60}$) around the $z$ axis.
Moreover, $\bm{e}_{20} = (0.75, 0.67, 0)$, $\bm{e}_{30} = (0.75, -0.67, 0)$.  
$\bm{e}_{40}$ ($\bm{e}_{50}$) is obtained by $\pi$ rotation of $\bm{e}_{20}$ ($\bm{e}_{30}$).

The polarization vectors $\bm{e}_{k2}$ in the second term at the right-hand-side (rhs) of Eq. (\ref{3qmsdw1}) are given by $\bm{e}_{12} = \bm{e}_{72} = (-0.96, -0.28, 0)$, $\bm{e}_{62} = \bm{e}_{82} = (-0.96, 0.28, 0)$, $\bm{e}_{22} = \bm{e}_{42} = (0.45, -0.89, 0)$, and $\bm{e}_{32} = \bm{e}_{52} = (0.45, 0.89, 0)$.
The site-dependent phases are given by $\alpha_{k}=\pi/4$ for odd $k$ and $\alpha_{k}=-\pi/4$ for even $k$.  The polarization vectors in the last term at the rhs of Eq. (\ref{3qmsdw1}) are given by 
$\bm{e}_{14} = (-0.49, 0.87, 0)$, $\bm{e}_{64} = (0.49, 0.87, 0)$.  
$\bm{e}_{74}$ ($\bm{e}_{84}$) is obtained by $\pi$ rotation of $\bm{e}_{14}$ ($\bm{e}_{64}$).
Moreover, $\bm{e}_{24} = (-0.87, -0.49, 0)$, $\bm{e}_{34} = (0.87, -0.49, 0)$.  
$\bm{e}_{44}$ ($\bm{e}_{54}$) is obtained by $\pi$ rotation of $\bm{e}_{24}$ ($\bm{e}_{34}$) 
around the $z$ axis.

\begin{figure}[htbp]
\begin{center}
\vspace{2cm}
\includegraphics[width=10cm]{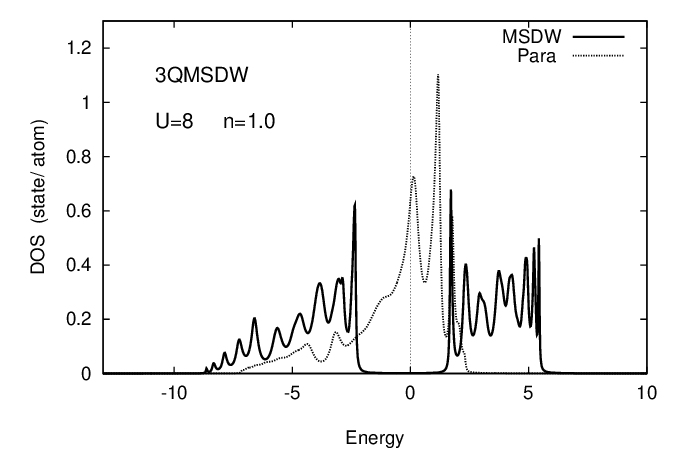}
\end{center}
\vspace{-2cm}
\caption{
DOS at $(n, U) = (1.0, 8)$. Solid curve: DOS in the $3Q$ Multiple Spin Density Waves (3QMSDW) state, 
dotted curve: DOS in the paramagnetic state. 
}
\label{dos108}
\end{figure}

The magnetic moments on site II $(k = 9, 14, 16, 19)$ at $(n, U) = (1.0, 8)$ form a simple helical structure (H) such that
\begin{eqnarray}
\bm{m}(\bm{R}_{lk}) &=& m_{4} ( \bm{i} \cos \bm{Q}_{2} \cdot \bm{R}_{lk} 
+ \bm{j} \sin \bm{Q}_{2} \cdot \bm{R}_{lk} ) \ ,
\label{3qmsdw2}
\end{eqnarray}
where $m_{4} = 0.94$, $\bm{i}$ ($\bm{j}$) is the unit vector along the $x$ ($y$) axis.
The magnetic moments on site II $(k = 10 \sim 13, 15, 17, 18, 20)$ consist of the site-dependent $\bm{Q}_{0}$ polarization, helical SDW, and the TSDW (H+TSDW+$Q_{0}$):
\begin{eqnarray}
\bm{m}(\bm{R}_{lk}) &=& m_{5} \, \bm{e}_{k0}  \nonumber \\
&+& m_{6} \, [ \bm{i} \cos (\bm{Q}_{2} \cdot \bm{R}_{lk} + \beta_{k}) 
+ \bm{j} \sin (\bm{Q}_{2} \cdot \bm{R}_{lk} + \beta_{k}) ]  \nonumber \\
&+& m_{7} \, \bm{e}_{k4} \cos ( \bm{Q}_{4} \cdot \bm{R}_{lk} + \gamma_{k}) \ .
\label{3qmsdw3}
\end{eqnarray}
Here $m_{5}=0.24$, $m_{6}=0.86$, $m_{7}=0.43$, 
$\bm{e}_{100}=-\bm{j}$, and $\bm{e}_{150}=-\bm{i}$.   
$\bm{e}_{110}$, $\bm{e}_{120}$, and $\bm{e}_{130}$ ($\bm{e}_{170}$, $\bm{e}_{180}$, and $\bm{e}_{200}$) are obtained by successive $\pi/2$ rotations of $\bm{e}_{100}$ ($\bm{e}_{150}$) around the $z$ axis. 
The site-dependent phases in the second term are given by $\beta_{k}=\pi/2$ $(k = 11, 12, 15, 17)$ and $\beta_{k}=-\pi/2$ $(k = 10, 13, 18, 20)$.
The polarization vectors in the third term at the rhs are given by $\bm{e}_{104} = (0.75, 0.66, 0)$, $\bm{e}_{114} = (0.66, 0.75, 0)$. 
$\bm{e}_{134}$, $\bm{e}_{204}$, and $\bm{e}_{184}$ ($\bm{e}_{174}$, $\bm{e}_{154}$, and $\bm{e}_{124}$) are obtained by successive $\pi/2$ rotations of $\bm{e}_{104}$ ($\bm{e}_{114}$). 
The site-dependent phases are given by $\gamma_{k}=\pi/4$ $(k = 10, 13, 18, 20)$ and $\gamma_{k}=-\pi/4$ $(k = 11, 12, 15, 17)$. 
We note that the 3QMSDW at $(n, U) = (1.0, 8)$ is stabilized by a gap formation in the DOS at the Fermi level as seen from Fig. \ref{dos108}.
%
%
%
%

The 3QMSDW at $(n, U) = (0.9, 8)$ consists of the same wave vectors as those of $(n, U) = (1.0, 8)$.  The magnetic structure shows the same form as that of $(n, U) = (1.0, 8)$, but the helical structure (H) on site II $(k = 9, 14, 16, 19)$ changes to the superposition of the site-dependent $2Q$ multiple helical SDW and $\bm{Q}_{0}$ polarization (2QH+$Q_{0}$).

The 3QMSDW at $(n, U) = (0.8, 8)$ consists of the wave vectors $\bm{Q}_{1} = (0, 0, 0.1)$, $\bm{Q}_{3} = (0, 0, 0.3)$, $\bm{Q}_{5} = (0, 0, 0.5)$, and  forms a site-dependent 3QTSDW on the $xy$ plane. 
\begin{eqnarray}
\bm{m}(\bm{R}_{lk}) &=& 
m_{k1} \, \bm{e}_{k1} \cos \bm{Q}_{1} \cdot \bm{R}_{lk} 
+ m_{k2} \, \bm{e}_{k2} \sin \bm{Q}_{1} \cdot \bm{R}_{lk}  \nonumber \\  
&+& m_{k3} \, \bm{e}_{k3} \cos \bm{Q}_{3} \cdot \bm{R}_{lk} 
+ m_{k4} \, \bm{e}_{k4} \sin \bm{Q}_{3} \cdot \bm{R}_{lk}  \nonumber \\  
&+& \bm{e}_{k5} \, ( m_{k5} \cos \bm{Q}_{5} \cdot \bm{R}_{lk} 
+ sgn \, m_{k6} \sin \bm{Q}_{5} \cdot \bm{R}_{lk} )  \ .  
\label{3qmsdw4}
\end{eqnarray}
Here $sgn=1$ for sites I $(k = 1, 3, 4, 8)$ and sites II $(k = 10, 14 \sim 18)$, and $sgn=-1$ for the other sites of $k$.  $\bm{e}_{ki} \ (i=1 \sim 5)$ are the polarization vectors on the $xy$ plane.
The amplitudes $m_{kj} \ (j=1 \sim 6)$ in the three terms at the rhs are comparable to each other on site I, while the first term with $\bm{Q}_{1}$ is dominant on site II.

The 3QMSDW at $(n, U) = (0.9, 7)$ and $(0.9, 6)$ show the same type of 3QTSDW as that of $(n, U) = (0.8, 8)$.  
At $(n, U)=(0.9, 7)$, the 3QTSDW on site I $(k = 2, 3, 4, 5)$, however, changes to the superposition of  the helical state with $\bm{Q}_{1}$ and 2QTSDW (H+2QTSDW), and the TSDW with the wave vector $\bm{Q}_{5}$ ($\bm{Q}_{3}$) becomes the main term on site I (site II).
In the case of $(n, U) = (0.9, 6)$, the 3QTSDW on site I $(k = 1 \sim 8)$ changes to the site-dependent H+2QTSDW, while the site II $(k = 9 \sim 20)$ remain the 3QTSDW.

Finally, the 3QMSDW at $(n, U) = (1.0, 5)$ consists of the wave vectors  $\bm{Q}_{0} = (0, 0, 0)$, $\bm{Q}_{2} = (0, 0, 0.2)$, and $\bm{Q}_{4} = (0, 0, 0.4)$.
The magnetic moments on site I are given by the superposition of the helical state with $\bm{Q}_{4}$, the TSDW with $\bm{Q}_{2}$, and $\bm{Q}_{0}$ polarization state (H+TSDW+Q${}_{0}$). 
The magnetic moments on site II $(k = 9, 14, 16, 19)$ also show the H+TSDW+Q${}_{0}$ structure, but the wave vector of the helical (TSDW) state is $\bm{Q}_{2}$ ($\bm{Q}_{4}$)  instead of $\bm{Q}_{4}$ ($\bm{Q}_{2}$).  
The magnetic moments on site II $(k = 10 \sim 13, 15, 17, 18, 20)$ are given by the double $Q$ helical states and the $\bm{Q}_{0}$ polarization (2QH+Q${}_{0}$).  
As in the case of $(n, U) = (1.0, 8)$, the 3QMSDW at $(n, U) = (1.0, 5)$ is stabilized by the formation of the gap at the Fermi level in the DOS.


Calculated magnetic phase diagram is consistent with the physical picture obtained by the experimental data on the skyrmion (Sk) systems.
Note that there exist the DM interactions in the $\beta$-Mn type lattice system without inversion symmetry.  
When we take into account the DM interactions, 
the ferromagnetic region in the present calculations may be regarded as the region for the Sk formation due to the competition between the ferromagnetic interactions and the DM interactions (: F+DM type Sk region), while the helical region may be regarded as the region of the Sk formation due to the competition between the ferro- and antiferro-magnetic interactions (: F+AF type Sk region) because the exchange type antiferromagnetic interactions are usually stronger than the DM ones.

In order to clarify the relation between the present results and the experimental data, we adopt here the five-fold equivalent bands model~\cite{hase72,hub81} for transition metals with the $d$ electron number $n_{d}$.  The model is known to be useful for understanding the systematic change of magnetism with band filling~\cite{mori85,kake13}, though the orbital dependence of the DOS is not necessarily so small in general.
Then the electron number $n$ in our phase diagram may be regarded as $n \sim n_{d}/5$.  Note that $n=1$ corresponds to $n_{d} = 5$, and $n=2$ to $n_{d} = 10$ in this model.
The effective Coulomb interactions $U$ in the 3$d$ transition metals are roughly estimated to be half the band width~\cite{kake17}, $U \sim W/2$.  This means $U/|t_{1}| \sim 8$ for the face-centered cubic structure or body-centered cubic structure model with the nearest neighbor transfer integrals $t_{1}$.
 
Although the real Sk systems on the $\beta$-Mn type lattice contain the 2nd and 3rd elements, we assume here that the magnetic interactions are determined by the 3$d$ electron number of the systems.
Co${}_{10}$Zn${}_{10}$ system is then characterized by $n_{d} \sim 8$ and $U \sim 8$ in unit of $|t_{1}|$, {\it i.e.}, $(n, U) = (1.6, 8)$ in our model.  Therefore Co${}_{10}$Zn${}_{10}$ is the F+DM type Sk system according to our phase diagram (Fig. \ref{unpdncoll}).
Similarly, Fe${}_{2-x}$Pd${}_{x}$Mo${}_{3}$N is specified by $n_{d} \sim 7$ and $U \sim 8$, {\it i.e.}, $(n, U) = (1.4, 8)$ in our phase diagram. Thus, Fe${}_{2-x}$Pd${}_{x}$Mo${}_{3}$N belongs to the F+DM type Sk system in agreement with the picture suggested by the experiments.

On the other hand, the $d$ electron number of Co${}_{x}$Zn${}_{y}$Mn${}_{z}$ ($x+y+z=20$) changes from $n_{d} \sim 8$ to $n_{d} \sim 6$ with increasing $z$.  This means that $(n, U)$ change  from $(1.6, 8)$ to $(1.2, 8)$.
Therefore according to our phase diagram the Co${}_{x}$Zn${}_{y}$Mn${}_{z}$ system may change from the F+DM type Sk to the F+AF type Sk system with increasing $z$.
These results are in agreement with the physical picture of the Sk formation obtained from the experimental side~\cite{toku14,karu20,yu18,qian20}.

\section{Summary}

We have calculated the magnetic phase diagram of the Hubbard model on the $\beta$-Mn type lattice in the weak and intermediate Coulomb interaction regimes using the generalized Hartree-Fock approximation combined with the recursion method to clarify the magnetism 
of the $\beta$-Mn type transition-metal alloys and compounds, especially, magnetic interactions in the magnetic skyrmions (Sk). 
Starting from the ferro- and heli-magnetic spin configurations, we obtained the ferromagnetic state, the ferrimagnetic state, the helimagnetic state, the partially ordered SDW state, and the 3QMSDW states on the $U$-$n$ plane.

We found that the ferromagnetic state is stabilized in the region $n \gtrsim 1.5$ and $U \gtrsim 1/\rho_{\rm II}(0)$ in the $U$-$n$ phase diagram, while the ferrimagnetic state is stabilized  in the region $0.4 \lesssim n \lesssim 0.7$ and $U \gtrsim 4$.  
Here $\rho_{\rm II}(0)$ denotes the local DOS on site II at the Fermi level.  
Note that the magnetic moments on site I in the ferromagnetic state  
continuously change their sign around half filling with decreasing electron number $n$.  
In this sense, the ferromagnetic state is connected to the ferrimagnetic one.
 
The helimagnetic state is stabilized in the region between the ferri- and ferro-magnetic state: $0.7 \lesssim n \lesssim 1.5$ and $U \gtrsim 3$.  
We found that the wave number $Q$ does not exceed 0.5 in the $\beta$-Mn type lattice system. 
Instead, short-range antiferromagnetic correlations develop in a unit cell with decreasing electron number $n$, so that we obtained the helical states with antiferromagnetic local structure in a unit cell near half filling.  
We also found the $Q=0$ noncollinear magnetic structures in the same region.

In the self-consistent calculations, we obtained two kinds of complex magnetic structures, the partially-ordered SDW and the 3QMSDW.
We found the partially-ordered SDW state with $Q=0.5$ at $(n, U) = (1.4, 3)$ in the weak Coulomb interaction regime.  The magnetic moments on site I vanish there and the site II moments form a site-dependent noncollinear SDW.
We obtained the 3QMSDW consisting of the wave vectors $\bm{Q}_{0} = (0, 0, 0)$, $\bm{Q}_{2} = (0, 0, 0.2)$, and $\bm{Q}_{4} = (0, 0, 0.4)$ at $(n, U) = (1.0, 8)$, $(0.9, 8)$, and $(1.0, 5)$, respectively, while we obtained  
the 3QMSDW with $\bm{Q}_{1} = (0, 0, 0.1)$, $\bm{Q}_{3} = (0, 0, 0.3)$, and $\bm{Q}_{5} = (0, 0, 0.5)$ at $(n, U) = (0.8, 8)$, $(0.9, 7)$, and $(0.9, 6)$.  
These 3QMSDW show the site-dependent noncollinear magnetic structure on the $xy$ plane, and are stabilized by a gap formation in the middle of the DOS.

We have shown that calculated magnetic phase diagram is consistent with the magnetic interactions for the Sk in transition metal alloys and compounds on the $\beta$-Mn type lattice.
Taking into account the weak Dzyaloshinskii-Moriya (DM) interactions in the $\beta$-Mn type system, we can interpret the ferromagnetic region in our phase diagram as the region of the Sk formation due to competition between the ferromagnetic (F) and the DM interactions, {\it i.e.}, the F+DM type Sk region, and the helical region including the 3QMSDW as the region of the Sk formation due to competition between the ferro- (F) and the antiferro-magnetic (AF) interactions, {\it i.e.}, the F+AF type Sk region.  
On the basis of the phase diagram, we reached the conclusion that the Sk in the  
$\beta$-Mn type Co${}_{10}$Zn${}_{10}$ and Fe${}_{2-x}$Pd${}_{x}$Mo${}_{3}$N are the F+DM type, while the F+AF type Sk are possible in Co${}_{x}$Zn${}_{y}$Mn${}_{z}$ $(x+y+z=20)$ alloys, in agreement with the physical picture obtained from the experimental data~\cite{toku14,karu20,yu18,qian20}.

Although we calculated the phase diagram and clarified the magnetic interactions in the Sk systems on the $\beta$-Mn type lattice, we note that the present approach does not guarantee the global minimum of the total energy for the calculated structure. 
We have to make further self-consistent calculations taking into account other possible structures and applying advanced theories which automatically find the most stable structure~\cite{kake98,kimu00,uchida16} in order to make sure the present results.

We also note that most of the Sk lattices in the experimental data are found at finite temperatures.
In order to clarify the formation of the itinerant-electron Sk, we have to make finite-temperature calculations taking into account the DM interactions as well as spin fluctuations.
These calculations and theoretical investigations on the magnetic structure, we believe, will lead to deeper understanding of the formation of itinerant electron Sk in the $\beta$-Mn type system.







\end{document}